\documentclass[reprint, superscriptaddress, amsmath,amssymb, aps, pra]{revtex4-1}

\usepackage{graphics}
\usepackage{graphicx}
\usepackage{dcolumn}
\usepackage{bm}
\usepackage{amsmath}
\usepackage{amssymb}
\usepackage{xcolor}
\usepackage{dsfont}
\usepackage{braket}
\numberwithin{equation}{section}

\makeatletter
\renewcommand\paragraph{\@startsection{paragraph}{4}{\z@}%
  {-3.25ex \@plus -1ex \@minus -.2ex}%
  {1.5ex \@plus .2ex}%
  {\normalfont\normalsize\bfseries}*}
\makeatother

\begin{document}

\title{How Many Dark Neutrino Sectors Does Cosmology Allow?}

\author{Alan Zander}\email{alan.zander@tum.de}
\affiliation{Technical University Munich (TUM), James-Franck-Strasse 1, 85748 Garching, Germany}
 
\author{Manuel Ettengruber}\email{manuel@mpp.mpg.de}
\affiliation{Max-Planck-Institut f\"ur Physik (Werner-Heisenberg-Institut), F\"ohringer Ring 6, 80805 M\"unchen, Germany}

\author{Philipp Eller}\email{philipp.eller@tum.de}
\affiliation{Technical University Munich (TUM), James-Franck-Strasse 1, 85748 Garching, Germany}

\begin{abstract}
We present the very first constraints on the number of Standard Model (SM) copies with an additional Dirac right-handed neutrino. From cosmology, we are able to pose strong limits on large regions of the parameter space. Moreover, we show that it is possible to account for the right dark matter density in form of stable particles from the dark sectors.
\end{abstract}

\maketitle

\section{Introduction}
\label{sec:level1}
The mass of the Higgs boson is affected by quantum corrections, which lead to a quadratic divergence that, in the absence of new physics, would tend to push its mass up to the cut-off of the Standard Model (SM), at around the Planck mass $M_P = 1.22 \cdot 10^{19}$ GeV. Accounting for the discrepancy between the expected scale and the actual observed Higgs mass $M_{\text{Higgs}} \approx 125$ GeV \cite{particle2020review} represents one of the major challenges in particle physics and it is known as the \textit{hierarchy problem} \cite{dvali2017strong,koren2020hierarchy}. The more conventional approach relies on mechanisms that strive to explain the, otherwise unnatural, cancellation of terms necessary to make sense of the observed Higgs mass. However, there have been other attempts to address the hierarchy problem from a totally different perspective. This is the case for theories that assume a smaller fundamental scale of gravity, narrowing hereby the gap between the Higgs mass and the cut-off, or in other words, between the weak and gravity scale. The Planck scale is degraded to an effective gravity scale that results, for example, from the large size of extra dimensions \cite{arkani1998hierarchy,Arkani-Hamed:1998sfv} or the large amount of extra particle species \cite{dvali2010black,Dvali:2007wp}. In this work, we will focus on the latter approach of assuming "many species". More specifically, we will follow \cite{Dvali:2009ne} and assume many copies of the SM. 

In addition to taking care of the hierarchy problem, introducing many copies of the SM immediately generates very interesting dark matter (DM) candidates. See later, or \cite{dvali2009dark}, for realizations of DM within this framework. Yet another unresolved question that can be addressed is the smallness of the active neutrino masses. If those masses arise through the Higgs mechanism as it is thought to be the case for the charged leptons and quarks, then it is puzzling why neutrinos should have a mass many orders of magnitudes smaller than the rest of the fermions, in particular within the same generation. The most established course of action in this regard is the well-known Seesaw mechanism \cite{Minkowski:1977sc,Gell-Mann:1979vob,Yanagida:1980xy,Mohapatra:1979ia,Mohapatra:2004zh}, which imposes a Majorana nature on the neutrinos and implies the violation of lepton number conservation. It also requires the existence of very heavy right-handed neutrinos (RHNs), whose Majorana mass is usually linked with more new physics \cite{Weinberg:1979sa}. With the framework investigated, however, we can tackle the neutrino mass problem by considering Dirac RHNs and invoking, once again, a large amount of SM copies \cite{Dvali:2009ne,Ettengruber:2022pxf}. We see, therefore, that the motivation of our framework goes beyond the hierarchy problem, making it worthwhile to explore further, even independently of the hierarchy problem.

In this work, we will focus on the neutrino sector of the theory and its cosmological impact, which may allow us, for the first time, to constrain the number of SM copies. The paper is organized as follows. In Section (\ref{sec: model}), we will engage in more detail with the model in question. In Section (\ref{sec: production}) we address the cosmological production of RHNs, while in Section (\ref{sec: constraints}) we discuss their cosmological impact and we show our results. We then conclude in Section (\ref{sec: conslusions}).

\section{MODEL\label{sec: model}}
As mentioned before, we will follow here \cite{Dvali:2009ne} and assume $N$ copies of the SM, enlarged by a Dirac RHN.

It has been shown \cite{dvali2010black} that by introducing $N_{\text{PS}} \gg 1$ particle species, the fundamental scale of gravity $M_f$ must fulfill the following relation

\begin{equation}
M_f \leq \frac{M_P}{\sqrt{N_{\text{PS}}}}.
\label{eq: condition for Mf with large particle species N}
\end{equation}
 Note that since every SM copy consists of $N_{\text{SM}} = \mathcal{O}(100)$ particle species, we have $N_{\text{PS}} = N N_{\text{SM}}$. Equation (\ref{eq: condition for Mf with large particle species N}) makes evident that for a large number of SM copies $N \gg 1$, the fundamental scale of gravity $M_f$ is suppressed with respect to the Planck mass $M_P$, narrowing the gap between the electroweak (EW) and gravity scale. Only for $N \sim 10^{30}$ we completely solve the hierarchy problem, $M_f \lesssim 1 \ \text{TeV} = \mathcal{O}(M_{\text{Higgs}})$. Yet, \textit{any} number $N$ of SM copies contributes to the "softening" of the hierarchy problem, whose complete solution could consist of a combination of different mechanisms. 

\bigskip
So our model consists of $N$ copies of the SM extended by a Dirac RHN. For the sake of simplicity, we will only consider one flavor following also \cite{Dvali:2009ne}, where it was shown that the masses generated after EW symmetry breaking will be very small as a consequence of unitarity and a large number of copies. The corresponding Lagrangian then reads

\begin{equation}
    \mathcal{L}_{\text{Yukawa}} = \left( \overline{L} \epsilon \Phi^* \right)_i \lambda_{ij} \nu_{R, j} + \text{h.c.,}
    \label{eq: Yukawa coupling Lagrangian}
\end{equation}
where $\Phi_i$ and $L_i$ stand for the Higgs and lepton SU(2)-doublets of the $i^{\text{th}}$ SM copy while $\epsilon \equiv i \sigma_2$ is the totally antisymmetric SU(2) tensor ($\sigma_k$ refers to the $k^{\text{th}}$ Pauli-matrix). Here, $\lambda$ is a $N \times N$ Yukawa matrix in the space of copies and $\nu_{R, j}$ is the RHN of the $j^{\text{th}}$ SM copy. The vertex of equation (\ref{eq: Yukawa coupling Lagrangian}) is only possible because RHNs are SM-singlets and can communicate beyond their own copy. However, it is not the only renormalizable and gauge-invariant interaction mixing sectors, since the Higgs portal and the photon kinetic mixing are also allowed. Nonetheless, these are more strongly suppressed than the neutrino Yukawa couplings \cite{Dvali:2009ne}. Even if they dominate the communication among copies, their influence will only strengthen the constraints discussed in this work. Our choice of not considering them is therefore very conservative and at the same time maintains the predictivity of the theory by not introducing new free parameters. 

Now back to our Yukawa interaction (\ref{eq: Yukawa coupling Lagrangian}). Assuming all SM copies interact the same way with the rest of copies (this is a consequence of the exact permutation symmetry introduced in \cite{Dvali:2009ne}), we arrive at the only possible configuration of the Yukawa matrix, namely

\begin{equation}
    \lambda =  \begin{pmatrix}
    a      & b      & \dots  & b \\
    b      & a      & \dots  & b \\
    \vdots & \vdots & \ddots & \vdots \\
    b      & b      & \dots  & a
    
    \end{pmatrix}  
    \label{eq: Yukawa matrix}.
\end{equation}
As already mentioned, the Lagrangian of equation (\ref{eq: Yukawa coupling Lagrangian}) generates a Dirac mass matrix $\mathcal{M} = \lambda \left\langle H_i \right\rangle$, which we can easily diagonalize in order to determine the mass eigenvalues and eigenstates. We obtain $(N - 1)$-degenerate states with mass

\begin{equation}
    m_\nu = (a - b) v,
    \label{eq: m_nu degenerate active neutrino mass}
\end{equation}
and one single heavier state with mass eigenvalue 

\begin{equation}
    m_H = \left(a - b + N b\right) v,
    \label{eq: m_H mass of heavy neutrino}
\end{equation}
where $ \left\langle H_i \right\rangle \equiv v = 174$ GeV is the Higgs vacuum expectation value in all copies. The latter mass eigenstate might be very heavy, given that its mass scales with the number of copies $N \gg 1$.

Both $a$ and $b$ in equation (\ref{eq: Yukawa matrix}) were introduced as Yukawa couplings. Since they are of the same nature, we do not expect them to be at totally different orders of magnitude, as was argued in \cite{Dvali:2009ne}. That is, the ratio $a/b$ should not be too large $a/b \not \gg 1$, which corresponds to the RHNs not being localized in one copy, as we would expect from singlets. In \cite{Dvali:2009ne} was also mentioned that due to unitarity, 

\begin{equation}
    b \leq \frac{1}{\sqrt{N}},
    \label{eq: unitarity condition}
\end{equation}
making the neutrino mass (\ref{eq: m_nu degenerate active neutrino mass}) small, for large $N$. For the sake of concreteness, we will assume $m_\nu = 0.1$ eV throughout the paper. Note that the philosophy in solving the neutrino mass problem within our model is very different from the Seesaw mechanism. The former gives an \textit{infrared} solution by introducing many light states, while the latter introduces one or few heavy states and it is therefore an \textit{ultraviolet} solution. 

Moreover, a key aspect of our model relies on the fact that we introduce, besides the RHN with its Yukawa couplings, \textit{only one} degree of freedom (the number of SM copies $N$) to approach the hierarchy problem and at the same time the neutrino mass problem (by suppressing $b \leq \frac{1}{\sqrt{N}}$) and even potentially DM.

\bigskip
Now, we can express the flavor eigenstates $\nu_i$, $1 \leq i \leq N$, as linear combinations of mass eigenstates. Without loss of generality, we label our copy $i = 1$ and choose a mass basis, in which:

\begin{equation}
\begin{split} 
    \nu_1 & = \sqrt{\frac{N - 1}{N}} \nu_1^m + \frac{1}{\sqrt{N}} \nu_H^m, \\
    \nu_j & = \sum_{k=2}^{N-1} a_k^j \nu_k^m - \frac{1}{\sqrt{N\left(N-1\right)}} \nu_1^m + \frac{1}{\sqrt{N}} \nu_H^m.
\end{split}
\label{eq: flavor neutrino in term of mass eig}
\end{equation}
Here $\nu_j$ ($2 \leq j \leq N$) correspond to the flavor eigenstates of other copies and are therefore sterile neutrinos (SNs) from our perspective. $\nu_1^m$, $\nu_k^m$ ($2 \leq k \leq N-1$) correspond to the degenerated states with mass $m_\nu$, while $\nu_H^m = \frac{1}{\sqrt{N}} \sum_{i=1}^N \nu_i$ is the heavier state with mass $m_H$. We call $\nu_k^m$  $\nu_H^m$ sterile-like mass eigenstates or simply SNs, when understood from the context that these are not flavor states. The coefficients $a_k^j$ in (\ref{eq: flavor neutrino in term of mass eig}) must satisfy some conditions due to orthonormality

\begin{equation}
\begin{split}
    & \sum_{k=2}^{N-1} a_k^j a_k^{j\prime} = \delta_{j j\prime} - \frac{1}{N-1} \\
    & \sum_{j=2}^{N} a_k^j a_{k\prime}^{j} = \delta_{k k\prime} \\
    & \sum_{j=2}^{N} a_k^j = 0 \ (2 \leq k \leq N-1).
\end{split}
\label{eq: unitarity conditions for ajk}
\end{equation}
From equation (\ref{eq: flavor neutrino in term of mass eig}) we see that the heavier state $\nu_H^m$ interacts to the same extent with all copies via its mixing angle

\begin{equation}
    \text{sin} \theta = \frac{1}{\sqrt{N}},
    \label{eq: mixing angle of heavy state}
\end{equation}
which is determined by the number of copies $N$. This is one of the reasons behind the model's high predictability. Note that this mixing vanishes for increasing $N \rightarrow \infty$, recovering the no-new physics scenario in the SM neutrino sector.

We can now easily compute the oscillation probabilities, provided coherence is not lost and $\nu_H^m$ is energetically accessible,

\begin{equation}
\begin{split}
    P_\text{surv} & \equiv P(\nu_1 \rightarrow \nu_1) = 1 - 4 \frac{N-1}{N^2} \text{sin}^2\left(\frac{\Delta m^2 L}{4 E}\right) \\
    P_j & \equiv P(\nu_1 \rightarrow \nu_j) = \frac{4}{N^2}\text{sin}^2\left(\frac{\Delta m^2 L}{4 E}\right),
\end{split}
\label{eq: osc prob}
\end{equation}
where $\Delta m^2 \equiv m_H^2 - m_\nu^2$.

\section{PRODUCTION IN THE EARLY UNIVERSE \label{sec: production}}

There are mainly two approaches when it comes to the production of both light ($\nu_j^m$, $j\geq 2$) and heavy ($\nu_H^m$) SNs in the early universe. Either they achieve equilibrium at some point in the history of the universe, when they permanently interact with the primordial plasma until they eventually decouple from the thermal bath (\textit{Freeze-Out}). Or, they are always out of equilibrium and are only produced through inelastic processes without ever interacting with the primordial plasma (\textit{Freeze-In}). We point out that only our SM copy can be present in the early universe (at least to the same extent), otherwise we would violate several cosmological constraints, some of which will be addressed further below. Therefore, we will always presume a primordial thermal bath composed exclusively of particles of our copy. In \cite{dvali2009dark}, for instance, inflationary and reheating mechanisms were proposed to achieve this very naturally. The most efficient production mechanisms arise from their only direct interactions via the Yukawa coupling (\ref{eq: Yukawa coupling Lagrangian}) and the mixing angle (\ref{eq: mixing angle of heavy state}).

\subsection{Freeze-Out}

If in equilibrium, SNs are simply Fermi-Dirac distributed. As the universe cools down, the annihilation rate will decrease until it falls below the expansion rate of the universe, such that the effective annihilation rate vanishes and the comoving number density remains constant, unless SNs are unstable. If SNs are relativistic, or at least they were at the time of decoupling from the thermal bath, their distribution function reads

\begin{equation}
    f_F(E) = \frac{1}{e^{\frac{E}{T}} + 1},
    \label{eq: Fermi distribution function}
\end{equation}
with $E$ the energy and $T$ the temperature of the SNs, which in general may differ from the temperature of the thermal bath. The number density, defined generally as 

\begin{equation}
    n(T) \equiv g \int \frac{d^3 \vec{p}}{(2 \pi)^3} f(E, T),
\label{eq: def number density}
\end{equation}
where $g$ stands for the \textit{internal} degrees of freedom, has the following form in equilibrium:

\begin{equation}
    n_F(T) = \frac{3 \zeta[3]}{2 \pi^2} \left(\frac{g_F}{2}\right) T^3.
\label{eq: Fermi-Dirac number density}
\end{equation}
 If SNs are non-relativistic at decoupling, their relic density will be exponentially suppressed and can be neglected.

\subsection{Freeze-In}

SNs that never achieve equilibrium are usually assumed to be absent in the very early universe. However, they can be populated through inelastic processes like Higgs decays, inverse decays, and oscillations of active neutrinos that are constantly produced and destroyed in the primordial plasma. This usually happens at high temperatures, such that at some point the production effectively ceases and again, the comoving particle density becomes constant. In the following, we will study the dominant production mechanisms when out of equilibrium of both light and heavy SNs, that as we mentioned, emerge from the Yukawa coupling and the mixing angle. We will describe separately the production for $m_H < M_\text{Higgs}$ and $m_H > M_\text{Higgs}$:

\paragraph{Regime: $\mathbf{m_H} \boldsymbol{<} \mathbf{M}_\textbf{Higgs}$}
\subsubsection{Active Neutrino Oscillations}

SNs can be non-resonantly produced through oscillations induced via incoherent interactions of the active neutrinos with the thermal plasma \cite{dodelson1994sterile}. In order to determine the distribution function $f_j$ of $\nu_j$, we resort to the Zeno-ansatz approach \cite{kishimoto2008lepton,abazajian2003cosmological}, neglecting the inverse process proportional to $f_j \ll f_1$,

\begin{equation}
    \left(\frac{\partial}{\partial t} - H p \frac{\partial}{\partial p}\right) f_j(p, t) = \Gamma^j_{\text{conv}} f_1(p, t) 
    \label{eq: Boltzmann kinetic eq.}
\end{equation}
where $H$ is the Hubble expansion rate of the universe, $\Gamma^j_{\text{conv}} = \frac{\Gamma_a}{2} \left\langle P_j^T \right\rangle$ the effective conversion rate of $\nu_1$ into $\nu_j$, $\Gamma_a$ is the interaction rate of active neutrinos with the plasma \cite{cline1992constraints}, while $ \left\langle P_j^T \right\rangle$ is the averaged transition probability $P_j$ at finite temperature. We can then solve equation (\ref{eq: Boltzmann kinetic eq.}) also for the number density $n_j$ analytically, assuming no lepton asymmetries, namely (see Appendix \ref{sec: osc prob})

\begin{equation}
    n_j(T) \approx 1.27 \cdot 10^2  \sqrt{\frac{10.8}{g_*(T_{\text{max}})}} \left(\frac{\sqrt{\Delta m^2}}{\text{eV}}\right) \left(\frac{T^3}{N^2}\right) ,
    \label{eq: nj from DW mechanism}
\end{equation}
where $g_*$ is the number of relativistic degrees of freedom and the temperature $T_{\text{max}} \approx 13.3 \ \text{MeV} \left(\frac{m_H}{\text{eV}}\right)^{\frac{1}{3}}$ corresponds to the maximal production rate. These sterile flavor states propagate and quickly decohere into heavy and light mass eigenstates due to their different group velocities \cite{farzan2008coherence,long2014detecting}. The wave packages of the light neutrinos $\nu_i^m (1 \leq i \leq N-1)$  do not come apart because of their degeneracy, but they constitute as a whole a mass eigenstate $\Tilde{\nu}_j^m \equiv \sqrt{\frac{N}{N-1}}\left[\sum_{k=2}^{N-1} a_k^j \nu_k^m - \frac{1}{\sqrt{N(N-1)}} \nu_1^m\right]$ of mass $m_\nu$. Hence, $\nu_j = \frac{1}{\sqrt{N}} \nu_H^m + \sqrt{\frac{N-1}{N}} \Tilde{\nu}_j^m$, for all $j=2, \dots, N$. The number densities of both light and heavy mass states are then a weighted sum over the flavor states \cite{weiler1999relic,fuller2009quantum},

\begin{equation}
\begin{split}
    & n_\ell \equiv \sum_{j=2}^{N} \left|\sqrt{\frac{N-1}{N}}\right|^2 n_j = \frac{\left(N-1\right)^2}{N} n_j, \\
    & n_H = \sum_{j=2}^{N} \left|\frac{1}{\sqrt{N}}\right|^2 n_j = \frac{N-1}{N} n_j,
\end{split}
\end{equation}

where $n_j(T)$ is given in equation (\ref{eq: nj from DW mechanism}). Of course, we have $n_\ell + n_H = \sum_j n_j$, and we approximate $n_\ell \approx N n_j$ and $n_H \approx n_j$. After production, heavy SNs eventually start decaying, such that the number density $n_H$ becomes exponentially suppressed.

For very heavy SNs with $m_H \gtrsim \mathcal{O}(M_\text{Higgs})$, we have $T_{\text{max}} \lesssim m_H$. This means that $\nu_H^m$ is non-relativistic or even energetically inaccessible at the time of production and our equations break down.

\subsubsection{Higgs Decays}
 Via (\ref{eq: Yukawa coupling Lagrangian}), our Higgs $H_1$ might decay into SNs. We write the Lagrangian in mass eigenstates and keep only the relevant terms,

 \begin{equation}
    \mathcal{L}_\text{Yukawa}^\text{BSM} \supset \sqrt{N-1} b \overline{\nu_1^m} H_1 \nu_{H, R}^m + \text{h.c.}
\end{equation}
Our Higgs does not couple to $\nu_k^m$ ($2 \leq k \leq N-1$) due to the conditions (\ref{eq: unitarity conditions for ajk}). Hence, in the early universe, the Higgs $H_1$ of our copy eventually decays into $\nu_1^m$ and $\nu_H^m$ with the respective probability. The active-like neutrino $\nu_1^m$ quickly equilibrates, while $\nu_H^m$ is populated out of equilibrium. The production occurs at temperatures comparable to the mass of the decaying particle, $T_\text{prod} \sim M_{\text{Higgs}}$, before $H_1$ disappears permanently from the thermal bath. We note at this point that other decay channels into SNs from our SM are mediated by the mixing angle $\theta$ from equation (\ref{eq: mixing angle of heavy state}) and are extremely suppressed at high temperatures (see again Appendix \ref{sec: osc prob}). This also means $\nu_1 \approx \nu_1^m$. Then, the heavy SN yield $Y_H^\infty \equiv \frac{n_H}{s}$ ($n_H$ is the number density of the heavy SNs and $s$ is the entropy density), after production via Higgs decays has ceased, can be approximated by \cite{hall2010freeze} (see Appendix \ref{sec: Production Decays})

\begin{equation}
    Y_H^\infty \approx \frac{135}{8 \pi^3 (1.66) \sqrt{g_*} g_*^S} \left(\frac{M_P \Gamma_{H_1 \rightarrow \nu_H \nu_1}}{M_{\text{Higgs}}^2}\right),
    \label{eq: number density from decays}
\end{equation}
where $\Gamma_{H_1 \rightarrow \nu_H \nu_1} = \frac{(N-1) b^2}{16 \pi} M_{\text{Higgs}} \left(1 - \frac{m_H^2}{M_{\text{Higgs}}^2}\right)^2$ is the decay rate for the process $H_1 \rightarrow \nu_1^m \nu_H^m$ and $g_*$, $g_*^S$ should be evaluated at $T = T_\text{prod}$. Further on, our heavy neutrino may decay itself, mainly into dark sectors. This state couples democratically to all copies, so only a negligible fraction $\frac{1}{N}$ of the final states will belong to our SM. The relevant decay channels depend on the heavy neutrino mass, but we do not need to keep track of all decay products. We can simply estimate that a fraction $\frac{1}{\mathcal{O}(1)}$ of the heavy SN's energy density $\rho_H(T) = \langle p \rangle n_H(T) \approx 2.45 T n_H(T)$ will end up as dark radiation at the time it might have a cosmological impact ($T \lesssim 1$ MeV). But also energetically accessible, stable, and massive particles like dark electrons, positrons and eventually (anti-)baryons may contribute to the DM density. Note that these would not annihilate as every copy is very diluted and out of equilibrium. Hence, if the yield of heavy SN after production is given by $Y_H^\infty = n_H^\infty/s$, we would end up with a matter density

\begin{equation}
    \Omega_\text{DM} h^2 = \frac{Y_H^\infty s_0 h^2}{\rho_c} \mu.
    \label{eq: DM density parametrized by mu}
\end{equation}
We have defined the "effective mass" of the decay products of $\nu_H^m$ as $\mu \equiv \sum_a \langle N_a \rangle m_a$, where the sum goes over all final stable states with mass $m_a$ at the end of the decay chain. The critical density and the entropy density, today, are denoted as $\rho_c$, $s_0$, respectively.

\paragraph{Regime: $\mathbf{m_H} \boldsymbol{>} \mathbf{M}_\textbf{Higgs}$} 

\subsubsection{Inverse Decay}

For very large SN masses $m_H > M_\text{Higgs}$, the process $H_1 \rightarrow \nu_1^m \nu_H^m$ is not energetically allowed anymore. However, the inverse reaction, the annihilation of an active neutrino and a Higgs into $\nu_H^m$, is viable. This will partially happen in the unbroken phase of our copy, but since the other sectors are so diluted, they remain unaffected through the EW phase transition in our sector. This gives modifications of order $\mathcal{O}\left(\frac{1}{N}\right)$, which we neglect. 

As detailed in Appendix \ref{sec: Production Decays}, the yield of $\nu_H^m$, after production has stopped, is

\begin{equation}
    Y_H^\infty \approx \frac{135 g_H M_P}{(1.66)8\pi^3 m_H^2} \frac{\Gamma_{H \rightarrow H_1 \nu_1}}{g_*^S \sqrt{g_*}} \left(\frac{2 \mathcal{C}(x_0)}{3 \pi}\right), 
    \label{eq: yield for mH inverse decay}
\end{equation}
where $\Gamma_{H \rightarrow H_1 \nu_1} = \frac{\left(N-1\right) b^2}{32 \pi} m_H \left(1 - \frac{M_\text{Higgs}^2}{mH^2}\right)^2$ and $\mathcal{C}(x_0) = \int_{x_0}^\infty dx x^3 K_1(x)$ is a function of the cut-off $x_0 = \frac{m_H}{T_0}$ of the theory. In our scenario, this is $T_0 = M_f = \frac{M_P}{\sqrt{N_\text{PS}}} \approx \frac{M_P}{10 \sqrt{N}}$. For $x_0 \lesssim 1$, we can approximate $\mathcal{C}(x_0) \approx \frac{3 \pi}{2}$ and we have an infrared process, independent of the cut-off. This is no longer true for $m_H > M_f$, since there was never enough energy to produce heavy SNs and $\mathcal{C}(x_0)$ quickly vanishes for $x_0 > 1$. In reality, heavy SN decays do not set in when production is finished. With such large widths due to the many possible channels, the decay happens while being produced. This is also true for $\nu_H^m$ decays in the last section. However, this does not change the picture since in the end we are interested in the energy and number densities deposited in the dark sectors. The difference with the last section is that here the decaying particle is not in equilibrium, but frozen-in. Assuming that $\nu_H^m$ only decays into $\sum_{j,k} H_j \nu_k^m$, with total width $\Gamma_H = \sum_{j,k} \Gamma_{\nu_H \rightarrow H_j \nu_k}$, we can estimate the yield of light SNs $Y_l^\infty = \sum_k Y_k^\infty$ as, 

\begin{equation}
    Y_l^\infty \approx \sum_{k} \frac{Y_H^\infty}{N} \approx Y_H^\infty,
\end{equation}
Moreover, $\nu_H^m$ decays while relativistic and we have $\rho_l^\infty \approx \frac{1}{2} \rho_H^\infty$. Furthermore, there will be also here contributions to the DM coming from the massive final states in the cascades initialized by the heavy SN. We can make the same approach as in equation (\ref{eq: DM density parametrized by mu}). Although, note that the "effective mass" $\mu$ of the final products is a function of the decaying particle's mass and therefore not the same in both cases.

\section{CONSTRAINTS \label{sec: constraints}}

The conceivable presence in the early universe of SNs, both light and heavy, might have a great impact on the evolution of the universe. This allows us to constrain the theory by determining the viable parameters that are in agreement with cosmological observables. We find that BBN and the flatness of the universe pose the strongest constraints and we will focus only on them. We then discuss our results, which are outlined in Figure \ref{fig: Plot Nvsmu_final}.

\subsection{BBN}
BBN is sensitive to the total energy density $\rho_\text{total}$ of the universe, to which also SNs contribute. In the radiation-dominated epoch, $\rho_\text{total} \approx \rho_R$, where $\rho_R$ is the radiation energy density. The latter is usually parametrized by the so-called \textit{effective number of neutrino species} $N_{\text{eff}} \equiv N_{\text{eff}}^{\text{SM}} + \Delta N_{\text{eff}}$, and is defined by

\begin{equation}
    \rho_{R} = \rho_\gamma \left(1 + \frac{7}{8}\left(\frac{4}{11}\right)^{4/3} N_{\text{eff}}\right),
    \label{eq: definition effective number of neutrinos}
\end{equation}
where $\rho_\gamma = \frac{\pi^2}{15} T_\gamma^4$ is the energy density of photons at temperature $T_\gamma$.
With no new physics we naively expect $N_{\text{eff}} = 3$ and $\Delta N_{\text{eff}} = 0$. However, there are some corrections due to the partial heating of neutrinos during $e^\pm$ annihilations, such that $N_{\text{eff}}^{\text{SM}} \approx 3.043 - 3.045$ \cite{de2016relic,mangano2002precision,cielo2023neff}.

So, in order not to change much the energy density of the universe, and hence the primordial nuclei abundances, we must impose $\Delta N_{\text{eff}} \lesssim 0.2$ \cite{aghanim2020planck}, at the time of BBN, $T_{\text{BBN}} \approx 1$ MeV. 

\subsubsection{Discussion}

Through the mechanisms discussed in Section \ref{sec: production}, both light and heavy SNs can be abundantly produced in the early universe. It turns out that heavy SNs will decay immediately after their production, prior to BBN, and therefore do not influence the cosmic expansion rate directly. However, they decay essentially into dark sectors, transferring their energy density. A considerable fraction $1/\mathcal{O}(1)$ will be in form of dark radiation at the time of BBN, contributing to $\Delta N_{\text{eff}}$. 

\begin{figure*}
 \centering
    \includegraphics[width=0.75\textwidth]{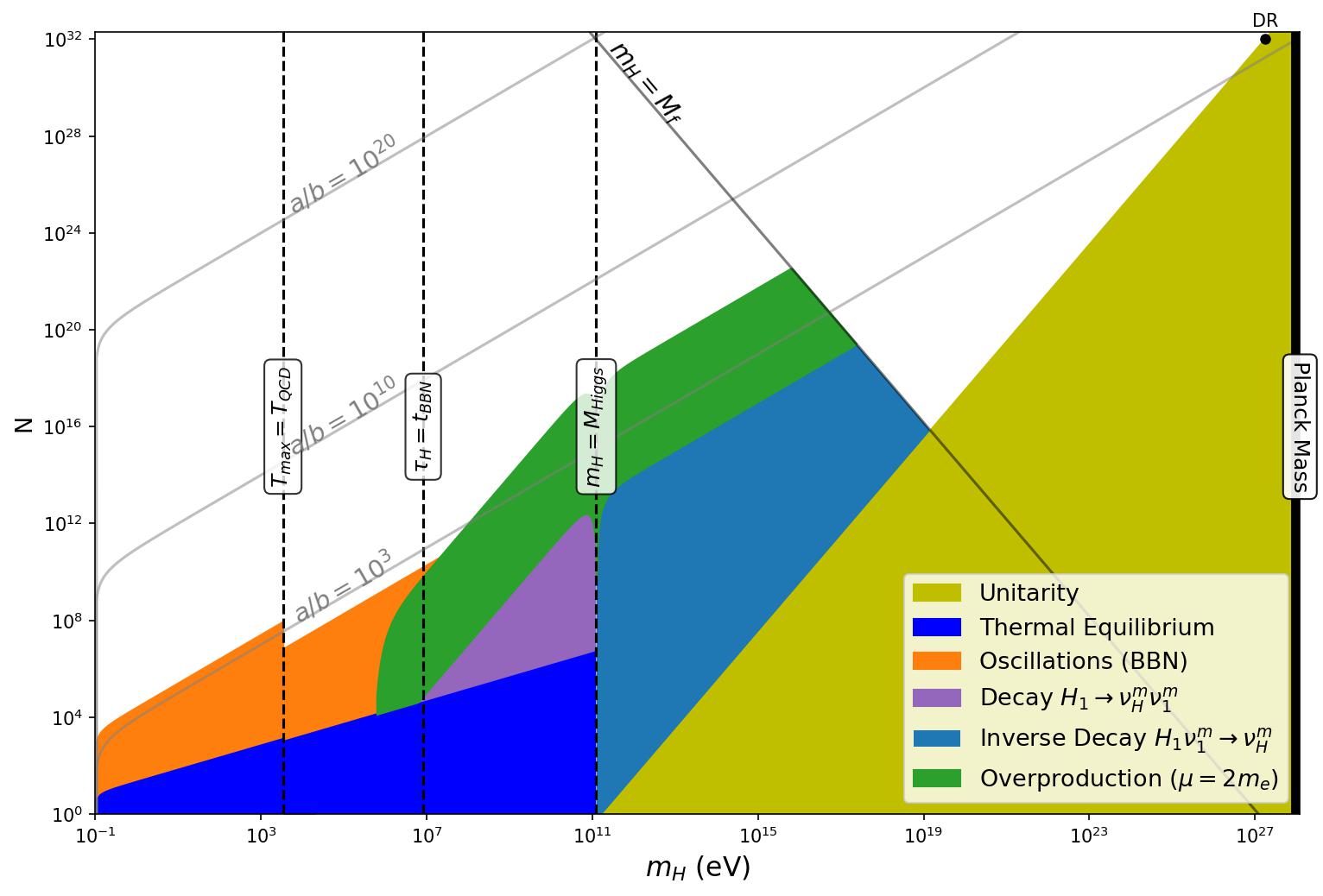}
   \caption{Cosmological constraints on the parameter space of our model spanned by the number $N$ of SM copies vs the mass $m_H$ of the heavy SN, where we have fixed the neutrino mass to be $m_\nu = 0.1$ eV. All constraints come from BBN, except for the green and olive areas which indicate overproduction and violation of unitarity, respectively. We have included regions corresponding to thermal equilibrium, non-resonant production via oscillations, production via decays $H_1 \rightarrow \nu_H^m \nu_1^m$ and inverse decays $H_1 \nu_1^m \rightarrow \nu_H^m$, as shown in the legends. Above the continuous black line, the heavy SN is heavier than the fundamental scale of gravity $m_H > M_f$ and its distance to the vertical Higgs line "$m_H = M_\text{Higgs}$", can be seen as to what extent the hierarchy problem is solved. The light gray lines reveal the ratio between the Yukawa couplings, $a/b = 10^3, 10^{10}, 10^{20}$. The point "DR" refers to the parameters used in \cite{dvali2009dark}.}
   \label{fig: Plot Nvsmu_final}
\end{figure*}

Light SNs, on the other hand, are still ultra-relativistic at BBN and will act as dark radiation. Parameters that allow light SNs to achieve equilibrium with our SM are excluded, since $\Delta N_{\text{eff}} = 1$ ($\Delta N_{\text{eff}} \approx N$, after summing over all copies!). In the left-hand side of Figure \ref{fig: Plot Nvsmu_final}, where $m_H < M_\text{Higgs}$, active neutrinos effectively oscillate into neutrinos in other copies, even if the latter are not in equilibrium, posing the strongest limit on the number of species. In this region, the temperature of maximal production $T_{\text{max}} \approx 13.3 \ \text{MeV} \left(\frac{m_H}{\text{eV}}\right)^{\frac{1}{3}}$ increases for larger masses $m_H$, such that active neutrino interact at a higher rate. Therefore, more SNs are produced, requiring larger $N$ not to spoil BBN. We have marked in Figure \ref{fig: Plot Nvsmu_final} the mass $m_H$ for which $T_{\text{max}}$ coincides with the QCD-temperature $T_\text{QCD}$, when the number of relativistic degrees of freedom $g_*$ changes abruptly. This generates a jump down in the SN relic density. In the right-hand side of Figure \ref{fig: Plot Nvsmu_final} we also obtain very strong limits on $N$ from BBN. Mediated always through the heavy SN, we produce light SNs (and dark photons) that might spoil BBN. However, the constraints are shadowed by the necessity of avoiding the overproduction of DM. Every time we produce a light SN, there is another heavy dark particle being created, like dark Higgs, $W$ or $Z$ bosons that can decay into their sectors and bend the universe (in contradiction to flatness). This will be addressed in the following.

\subsection{Flatness}
The total energy density \textit{today} equals the critical density $\rho_c = \frac{3 H_0^2}{8 \pi G} = 10.537 h^2$ GeV m$^{-3}$ \cite{belanger2018micromegas5}. Therefore, any contribution of the SNs or their decay products to the energy density, today, should not exceed the DM density $\Omega_{\text{DM}} h^2 \equiv \frac{\rho_\text{DM} h^2}{\rho_c} = 0.120 \pm 0.001$, otherwise it would be in contradiction with a flat universe \cite{aghanim2020planck}, as just mentioned. Note that one of the most interesting cases is when our scenario generates precisely the right density to explain DM.

\subsubsection{Discussion}

Matter density does not get red-shifted as radiation does, while the universe expands. For that reason, matter will dominate at some point the energy content of the universe. Any non-relativistic particle, whose energy density at the time of BBN was negligible, still has the chance of contributing substantially to the total density of the universe, in form of DM. This suggests that the massive final states outside our copy in the decay cascades (i.e. stable particles) of any particle, will contribute to the DM density, provided they do not disappear through annihilations. In our scenario, the only particle that decays into dark sectors and is produced in the early universe is $\nu_H^m$. Therefore if massive particles survive from these decays, they inevitably contribute to the DM density. This yields constraints even stronger than from BBN, predominantly in the right-hand side of Figure \ref{fig: Plot Nvsmu_final}. 

We have used equation (\ref{eq: DM density parametrized by mu}) to parametrize the contribution to DM. For $m_H \gtrsim 2 m_e \sim 1$ MeV upwards, heavy SNs mostly decay in such a way that at least one electron is produced at the end of the cascade, when all unstable particles have decayed. Of course, because of charge conservation, also a positron is produced. For the relevant parameters, each copy is so diluted that these $e^- e^+$-pairs (as well as proton-antiproton) do not annihilate into photons, as we have confirmed using the respective cross sections given in \cite{dvali2009dark,itzykson2012quantum}. This means that the "effective mass" $\mu \gtrsim 2 m_e \sim 1$ MeV is a very conservative lower limit, in particular for larger masses. Remarkably, we are able to reproduce the right DM density in form of electrons and positrons, as can be seen in Figure \ref{fig: Plot Nvsmu_final}. This corresponds to the edge of the green, overproduction region.

\subsection{Note}

We want to remark that our constraints, as usual for cosmological considerations, are model-dependent and may be circumvented, for example, by exploring cosmologies with reheating temperatures lower than the temperature at which SNs are most effectively produced. In contrast, we included the condition of equation (\ref{eq: unitarity condition}) required by unitarity. Furthermore, there are upper bounds on $N$ coming from axion physics that can complement our findings, and may be as strong as $N \lesssim 10^6$ \cite{ettengruber2023consequences}.

\section{CONCLUSIONS \label{sec: conslusions}}

We motivated at the beginning theories with many particle species. We focused in particular on the possible existence of many Standard Model copies, with an additional right-handed neutrino. Precisely this neutrino portal can trigger a considerable production in the early universe of the dark sectors that have a great impact on cosmology. We considered BBN and the flatness of the universe to pose very strong limits on the number of extra Standard Model copies, especially for heavy sterile neutrinos with very large masses. On the other hand, we found regions in the parameter space, where stable particles from other copies can account for the right dark matter density. 

\bigskip

\section*{Acknowledgments}
We would like to thank Gia Davli for the helpful discussions. This work has been supported by the Deutsche Forschungsgemeinschaft (DFG, German Research Foundation) under the Sonderforschungsbereich (Collaborative Research Center) SFB1258 ‘Neutrinos and Dark Matter in Astro- and Particle Physics’.

\appendix

\section{Non-Resonant Production via Oscillations \label{sec: osc prob}}

In vacuum, and in the mass basis, the Hamiltonian is

\begin{equation}
    \hat{H} = \begin{pmatrix}
    E_H     & 0      & \dots  & 0 \\
    0       & E      & \dots  & 0 \\
    \vdots  & \vdots & \ddots & \vdots \\
    0       & 0      & \dots  & E
    \end{pmatrix} = \text{diag}\left(\Delta, 0, \dots, 0\right) + E \mathds{1}_N
\label{eq: H in vacuum in mass eig},
\end{equation}
where $\mathds{1}_N$ is the $N \times N$ identity matrix, $E_H = \sqrt{m_H^2 + p^2}$, $E = \sqrt{m_\nu^2 + p^2}$ and $\Delta \equiv E_H - E$. In the interacting basis, the Hamiltonian reads 
\begin{equation}
    \hat{H} \rightarrow S \hat{H} S^T = \frac{\Delta}{N} \mathbf{1}_N+ E \mathds{1}_N,
\end{equation}
with $\mathbf{1}_N$ the $N \times N$ constant matrix, i.e with only ones in all entries. The orthonormal matrix $S$ diagonalizes the Dirac mass matrix $\mathcal{M} = \lambda v$ and can be chosen to be

\begin{equation}
   S =  \left(\begin{array}{cc|c}
    \frac{1}{\sqrt{N}} & \sqrt{\frac{N}{N-1}}     & 0, \dots, 0 \\ \hline
    \frac{1}{\sqrt{N}} & \frac{-1}{\sqrt{N(N-1)}} &   \\ 
    \frac{1}{\sqrt{N}} & \frac{-1}{\sqrt{N(N-1)}} &  \left(a_k^j\right) \\
    \vdots             & \vdots                   &   \\
    \frac{1}{\sqrt{N}} & \frac{-1}{\sqrt{N(N-1)}} &   
    \end{array} \right),
    \label{eq: Choice of matrix S, orthogonal}
\end{equation}
such that

\begin{equation}
    \nu^m = S^T \nu, 
\label{eq: relation between mass and interaction eigenstates}
\end{equation}
and 
\begin{equation}
\begin{split}  
    & \nu \equiv \nu_L + \nu_R = \left(\nu_1, \dots, \nu_N\right)^T, \\
    & \nu^m \equiv \nu_L^m + \nu_R^m= \left(\nu_H^m, \nu_1^m \dots, \nu_{N-1}^m\right)^T.
\end{split}
\label{eq: definition of mass and interaction eigenstate vectors}
\end{equation}
Due to orthonormality, the coefficients $a_k^j$ in (\ref{eq: Choice of matrix S, orthogonal}) must fulfill the conditions given in equation (\ref{eq: unitarity conditions for ajk}). Now, at finite temperature, the energy of the interacting neutrino of our SM, $\nu_1$, gets shifted by the finite density potential $V$,

\begin{equation}
    \hat{H}_T = \frac{\Delta}{N} \mathbf{1}_N+ E \mathds{1}_N - \text{diag}(V, 0, \dots, 0).
\end{equation}
Actually, the potential matrix should look like 

\begin{equation}
    \hat{V} =  \begin{pmatrix}
    V + V_a      & V_{ab}      & \dots  & V_{ab} \\
    V_{ab}      & 2V_b      & \dots  & V_b \\
    \vdots & \vdots & \ddots & \vdots \\
    V_{ab}      & V_b      & \dots  & 2V_b
    
    \end{pmatrix}  
    \label{eq: potential matrix},
\end{equation}
since for each pair $\nu_i$, $\nu_j$ there is a (or two, for $i=j$) bubble diagram(s) with a thermal propagator of $\nu_1$, whose background generates the potential \cite{notzold1988neutrono}. In the first row and column, there is also the contribution from a tadpole diagram. However, $V_a, V_{ab}, V_b$ are suppressed with respect to $V$ by $a^2, ab, b^2$, respectively. In particular, in the relevant region of the parameter space, we have $\frac{a}{b} \lesssim 10^3$ and $b < \frac{1}{N}$, such that even $N b^2 < \frac{1}{N}$ is suppressed. Hence, $\hat{V} \approx \text{diag}(V, 0, \dots, 0)$. 

We need to diagonalize $\hat{H}_T$ to find the mass eigenstates and energies at finite temperature. We find the eigenvalues 
\begin{equation}
    \begin{split}
        & E_H^T = \frac{1}{2}\left[\Delta + V + \sqrt{\Delta^2 + 2(N-2)\frac{\Delta V}{N} + V^2}\right] + E, \\
        & E_1^T = \frac{1}{2}\left[\Delta + V - \sqrt{\Delta^2 + 2(N-2)\frac{\Delta V}{N} + V^2}\right] + E,
    \end{split}
\end{equation}
while the rest of energies, and their respective eigenstates, remain unmodified. The eigenvectors (not normalized yet) corresponding to $E_H^T$, $E_1^T$ are 

\begin{equation}
    \begin{split}
        & \nu_H^{m, T} = \left(B_H, 1, \dots, 1\right)^T, \\
        & \nu_1^{m, T} = \left(B_1, -1, \dots, -1\right)^T,
    \end{split}
    \label{eq: nuH and nu1 matter eigenvectors}
\end{equation}
where

\begin{equation}
    \begin{split}
        & B_H = B_a - B_b, \\
        & B_1 = B_a + B_b,
    \end{split}
\end{equation}
with 

\begin{equation}
    \begin{split}
        & B_a \equiv \sqrt{\frac{N^2}{4} + N\frac{N-2}{2}\frac{V}{\Delta} + \frac{N^2}{4}\frac{V^2}{\Delta^2}}, \\
        & B_b \equiv \frac{N-2}{2} + \frac{N}{2}\frac{V}{\Delta}.
    \end{split}
\end{equation}
The norm squared of the vectors of (\ref{eq: nuH and nu1 matter eigenvectors}) are 

\begin{equation}
    \mathcal{N}_i^2 = B_i^2 + (N - 1),
\end{equation}
for $i \in \{1, H\}$, such that we normalize $\nu_i^{m, T} \rightarrow \frac{\nu_i^{m, T}}{\mathcal{N}_i}$. Then, 

\begin{equation}
    \nu_1 = \frac{B_1}{\mathcal{N}_1} \nu_1^{m, T} + \frac{B_H}{\mathcal{N}_H} \nu_H^{m, T}
    \label{eq: nu1 in terms of mass eigenstates in matter}
\end{equation}
We define the mixing angle in matter as the amplitude $\braket{\nu_H^{m, T}|\nu_1}$,

\begin{equation}
    \begin{split}
        & \text{sin} \theta_T = \frac{B_H}{\mathcal{N}_H}, \\
        & \text{cos} \theta_T = \frac{B_1}{\mathcal{N}_1}.
    \end{split}
    \label{eq: mixing angle at finite T}
\end{equation}
Indeed, the mixing angle (\ref{eq: mixing angle at finite T}) at finite temperature reduces to (\ref{eq: mixing angle of heavy state}), for $V = 0$. The remaining interacting states can be expressed as

\begin{equation}
    \nu_j = \sum_{k=2}^{N-1} a_k^j \nu_k^m - \frac{1}{\mathcal{N}_1} \nu_1^{m, T} + \frac{1}{\mathcal{N}_H} \nu_H^{m, T}.
    \label{eq: nuj in terms of mass eigenstates in matter}
\end{equation}
All unitarity conditions from equation (\ref{eq: unitarity conditions for ajk}) still hold. From equations (\ref{eq: nu1 in terms of mass eigenstates in matter}, \ref{eq: nuj in terms of mass eigenstates in matter}) it is easy to compute the transition probabilities in matter:

\begin{equation}
    P_j^T = \frac{1}{N-1} \text{sin}^2 2 \theta^T \text{sin}^2 \left(\frac{\Delta^T L}{2}\right),
\end{equation}
where $L$ is the distance traveled and $\Delta^T \equiv E_H^T - E_1^T = \Delta \sqrt{\text{sin}^2 2 \theta + \left(\text{cos} 2 \theta + \frac{V}{\Delta}\right)^2}$, which has the same exact form as for the usual active-sterile scenario, as well as the mixing angle in matter \cite{abazajian2017sterile}

\begin{equation}
    \text{sin}^2 2 \theta^T  = \frac{\text{sin}^2 2 \theta}{\text{sin}^2 2 \theta + \left(\text{cos} 2 \theta + \frac{V}{\Delta}\right)^2}.
\end{equation}
Since $V > 0$, there are no resonances, and the oscillation time is much shorter than the interaction time, such that we can average the oscillatory term, yielding

\begin{equation}
    \left\langle P_j^T \right\rangle = \frac{1}{2}\frac{1}{N-1} \text{sin}^2 2 \theta^T.
\end{equation}
Note that this also means that we are justified in neglecting the quantum-Zeno effect \cite{boyanovsky2007sterile}, which arises from the destruction of quantum phase due to constant interactions. We are now ready to solve equation (\ref{eq: Boltzmann kinetic eq.}). Assuming $g_* = \text{const.}$, we can write the left-hand side as

\begin{equation}
    \left(\frac{\partial}{\partial t} - H p \frac{\partial}{\partial p}\right) f_j(p, t) = H x \frac{\partial f_j(x, y)}{\partial x},
\end{equation}
where $x = \frac{\sqrt{\Delta m^2}}{T}$, $y = \frac{p}{T}$. This is a first-order differential equation that can be solved analytically, even when including the damping factor coming from the Zeno-quantum effect \cite{boyanovsky2007sterile}

\begin{equation}
\begin{split}
    & \left\langle P_j^T \right\rangle \rightarrow \frac{1}{2}\frac{1}{N-1} \frac{\text{sin}^2 2 \theta^T}{\left(\frac{D}{\Delta^T}\right)^2 + 1} = \\
    & \frac{1}{2}\frac{1}{N-1} \frac{\text{sin}^2 2 \theta}{\text{sin}^2 2 \theta + \left(\frac{D}{\Delta}\right)^2 + \left(\text{cos} 2 \theta + \frac{V}{\Delta}\right)^2}.
\end{split}
\end{equation}
Since active neutrinos are in equilibrium in the early universe, the \textit{decoherence} or \textit{damping function} $D$ reduces to $D = \frac{\Gamma_a}{2}$ \cite{bell1999relic}. At temperatures $T \ll M_W$ ($M_W$ being the mass of the $W$-boson) \cite{notzold1988neutrono}, 

\begin{equation}
    \Gamma_a (p, T)\simeq \chi_a \frac{7 \pi}{24} G_F^2 T^4 p,
    \label{eq: interaction rate active neutrinos two loops}
\end{equation}

with $\chi_e = \frac{13}{9}$ and $\chi_\mu = \chi_\tau = 1$. For concreteness, we will always use $a = e$. The \textit{weak potential} $V = V^L + V^T$ arise from neutrino neutral and charged current forward scattering on particles in the plasma \cite{abazajian2001sterile}. Assuming no asymmetries that generate lepton number, one has \cite{notzold1988neutrono,abazajian2001sterile} $V^L = 0$ and $V^T = \frac{28 \pi G_F^2 \text{sin}^2 \theta_W T^4 p}{45 \alpha} \left(\zeta_a + \frac{\text{cos}^2 \theta_W}{2}\right)$, where $\alpha$ is the fine-structure constant, $G_F$ the Fermi constant, $\theta_W$ the Weinberg angle and $\zeta_a (T) \approx \Theta \left(T - m_{l_a}\right)$ represents the contribution to the weak potential from the charged current, which depends on whether or not there is a background of charged leptons $l_a$ in the plasma at temperature $T$. 

Assuming no initial abundance, $f_j(x=0) = 0$ and integrating (\ref{eq: Boltzmann kinetic eq.}) up until $x \rightarrow \infty$, we obtain

\begin{equation}
\begin{split} 
    f_j(y) & = \frac{\chi_a}{\sqrt{W_a}} \cdot \frac{7 \pi^2 G_F M_P \sqrt{\Delta m^2}}{1.66 \sqrt{g_*(T_\text{max})}} \sqrt{\frac{45 \alpha}{56 \pi}} \frac{\text{sin}^2 2 \theta}{N-1} A f_1(y) \\
    & \approx \frac{\chi_a}{\sqrt{W_a}} \cdot 67.7 \sqrt{\frac{10.75}{g_*(T_\text{max})}} \frac{\sqrt{\Delta m^2}}{\text{eV}} \frac{\text{sin}^2 2 \theta}{N-1} f_1(y), 
\end{split}
\label{eq: distribution function non-resonant production appendix}
\end{equation}
where $W_a =\text{sin}^2 \theta_W \left(\zeta_a + \frac{\text{cos}^2 \theta_W}{2}\right)$ and

\begin{equation}
    A \equiv \sqrt{2}\frac{\sqrt{\sqrt{1 + \left(\frac{\chi_a 45 \alpha}{W_a 192}\right)^2} - \text{cos} 2 \theta}}{\sqrt{\text{sin}^2 2\theta + \left(\frac{\chi_a 45 \alpha}{W_a 192}\right)^2}} \approx 1
\end{equation}
is the result of the quantum Zeno effect. Note that the distribution of (\ref{eq: distribution function non-resonant production appendix}) is proportional to the equilibrium distribution $f_1(y) = f_F(y)$ ($f_F$ as in equation (\ref{eq: Fermi distribution function})). Integrating the distribution function $f_j$ of (\ref{eq: distribution function non-resonant production appendix}) over momentum as in (\ref{eq: def number density}) yields finally the number density $n_j$ of equation (\ref{eq: nj from DW mechanism}).


\section{Production via Decays and Inverse Decays \label{sec: Production Decays}}

Consider the process $1 \rightarrow a, b$, imposing $m_1 > m_a + m_b$. The Boltzmann equation describing the production of particle "$a$" is \cite{kolb1981early}

\begin{equation}
    H s x \frac{dY_a}{dx} = C_a,
\end{equation}
where $x\equiv \frac{m_1}{T}$, $Y_a \equiv \frac{n_a}{s}$ and $s = \frac{2 \pi^2}{45} g_*^S T^3$ is the entropy density with its corresponding number of degrees of freedom $g_*^S$. The collision term $C_a$, neglecting Pauli blocking and stimulated emission, as well as the inverse decay ($f_a, f_b \ll 1$) reads \cite{kolb1981early}

\begin{equation}
    C_a = \int d\Pi_1 d\Pi_a d\Pi_b  (2 \pi)^4 \delta^{(4)}(p_a + p_b - p_1) \cdot \\ |\mathcal{M}|^2 f_1 \ \text{.}
\end{equation}
where $d\Pi_i \equiv g_i \frac{d^3\Vec{p}_i}{\left(2 \pi\right)^3 2 E_i}$, $g_i$ is the number of internal degrees of freedom and $f_i$ correspond to the distribution function of $i \in \{1, a, b\}$. We can immediately recognize that the integration over $\vec{p}_a$ and $\vec{p}_b$ correspond to the decay rate of the process $1 \rightarrow a, b$, such that

\begin{equation}
    C_a = 2 m_1 \Gamma_{1 \rightarrow a, b} \int \text{d} \Pi_1 f_1(E_1) \ \text{,}
    \label{eq: Ca for arbitrary f1}
\end{equation}
which holds for an arbitrary distribution function $f_1$. In case the decaying particle is in equilibrium, we can approximate $f_1$ as Boltzmann-distributed $f_1(E_1) = e^{-\frac{E_1}{T}}$ and we obtain

\begin{equation}
C_a = \frac{m_1^3 g_1}{2\pi^2} \Gamma_{1 \rightarrow a, b} \cdot x^{-1} K_1(x),
\label{eq: collision term with K1}
\end{equation}
where 

\begin{equation}
    K_1 \left(x\right) \equiv x^{-1} \int_{x}^\infty du \sqrt{u^2 - x^2} e^{-u}
\end{equation}
is the modified Bessel function of the second kind of order one. Integrating (\ref{eq: collision term with K1}) over our "time"-parameter $x$ up to $x = \infty$ yields

\begin{equation}
    Y_a (\infty) \approx \frac{135 g_1 M_P}{(1.66)8\pi^3 m_1^2} \frac{\Gamma_{1 \rightarrow a, b}}{g_*^S \sqrt{g_*}} 
    \label{eq: yield for two particle decay}
\end{equation}

For the inverse process $a, b \rightarrow 1$, consider the different initial conditions of having vanishing small $f_1$ and Boltzmann-distributed $f_a$, $f_b$. This is a completely different scenario from before, note that $f_a$ was not Boltzmann-distributed, $f_1$ did not vanish and $f_b$ was arbitrary. The decay $1 \rightarrow a, b$ can be \textit{initially} neglected ($\propto f_1 \approx 0$). We obtain the following collision operator

\begin{equation}
    C_1 = 2 m_1 \Gamma_{1 \rightarrow a, b} \int d\Pi_1 e^{-\frac{E_1}{T}},
\end{equation}
which interestingly yields the same expression as in (\ref{eq: collision term with K1}), and therefore (\ref{eq: yield for two particle decay}).


%
\bibliographystyle{unsrt}
\bibliography{mybib}
%


\end{document}